\documentclass[12pt]{article}
\usepackage{graphicx, epsfig}
\begin{document}
\title{Vacuum Polarization of a Massless Scalar Field in the Background of a Global Monopole with Finite Core}
\author{J. Spinelly$^{1}$ {\thanks{E-mail: spinelly@fisica.ufpb.br}}  
and E. R. Bezerra de Mello$^{2}$ \thanks{E-mail: emello@fisica.ufpb.br}\\
1.Departamento de F\'{\i}sica-CCT\\
Universidade Estadual da Para\'{\i}ba\\
Juv\^encio Arruda S/N, C. Grande, PB\\
2.Departamento de F\'{\i}sica-CCEN\\
Universidade Federal da Para\'{\i}ba\\
58.059-970, J. Pessoa, PB\\
C. Postal 5.008\\
Brasil}

\maketitle

\begin{abstract}
In this paper we analyze the vacuum polarization effects of a massless scalar field in the background of a global monopole considering a inner structure to it. Specifically we investigate the effect of its structure on the vacuum expectation value of the square of the field operator, $\langle\hat{\Phi}^2(x)\rangle$, admitting a non-minimal coupling between the field with the geometry: $\xi {\cal{R}}\hat{\Phi}^2$. Also we calculate the corrections on the vacuum expectation value of the energy-momentum tensor, $\langle\hat{T}_{\mu\nu}\rangle$, due to the inner structure of the monopole. In order to develop these analysis, we calculate the Euclidean Green function associated with the system for points in the region outside the core. As  we shall see, for specific value of the coupling parameter $\xi$, the corrections caused by the inner structure of the monopole can provide relevant contributions on these vacuum polarizations.
\end{abstract}
\newpage

\section{Introduction}

A global monopole is a topological defect produced in the phase transition of a system composed of a self-coupling scalar field triplet, $\varphi^a$, whose original global $O(3)$ symmetry is spontaneously broken to $U(1)$. The scalar matter field plays the role of an order parameter which outside the the monopole's core acquire a nonvanishing value. The simplest model which gives rise to a global monopole is described by the Lagrangian density below, and has been analyzed by Barriola end Vilenkin \cite{BV}:  
\begin{equation}
{\mathcal{L}}=-\frac{1}{2}(\partial_{\mu}\varphi^{a})(\partial^{\mu}\varphi^{a})-\frac{\lambda}{4}(\varphi^{a}\varphi^{a}-\eta^{2})^{2} \ .
\label{La}
\end{equation}
The {\it ansatz} adopted to find monopole configuration is
\begin{equation}
\varphi^a(x)=\eta f(r)\frac{x^a}r \ ,
\end{equation}
with $x^ax^a=r^2$. 

The most general static metric tensor with spherical symmetry can be written as
\begin{equation}
\label{spherical}
ds^2=-B(r)dt^2+A(r)dr^2+r^2(d\theta^2+\sin^2\phi\  d\phi^2) \ .
\end{equation}
The Euler-Lagrange equation for the field $f(r)$ in the metric (\ref{spherical}) is
\begin{equation}
\frac{f''}{A}+\left[\frac{2}{Ar}+\frac1{2B}\left(\frac BA\right)'\right]f'-\frac{2f}{r^2}-\lambda\eta^2f(f^2-1)=0 \ , 
\label{EL}
\end{equation}
where the prime denotes differentiation with respect to $r$. The energy-momentum tensor associated with the matter field is given by
\begin{eqnarray}
T^t_t=-\eta^2\left[\frac{f^2}{r^2}+\frac{(f')^2}{2A}+\frac\lambda 4\eta^2(f^2-1)^2\right] \ , \nonumber \\ 
T^{r}_{r}=-\eta^2\left[\frac{f^2}{r^2}-\frac{(f')^2}{2A}+\frac{\lambda}{4}\eta^2(f^2-1)^2\right] \ , \\
T^\theta_\theta=T^\phi_\phi=-\eta^2\left[\frac{(f')^2}{2A}+\frac{\lambda}{4}\eta^2(f^2-1)^2\right] \ . \nonumber
\end{eqnarray}

The solutions to the scalar and gravitational fields can be obtained by using the above energy-momentum tensor as source of the Einstein equation, together with (\ref{EL}). However, unfortunately it is not possible to obtain exact analytical solutions to the complete set of non-linear coupled differential equations. Only numerical methods can provide solutions \cite{All}. \footnote{The numerical analysis of similar set of differential equations for the composite monopole system, i.e., a system composed by global and magnetic monopoles, have been developed in \cite{Mello1}.}

Analytical approximated solutions can be obtained in the region outside the monopole's core. It can be observed that in this region $f \sim 1$, and the energy-momentum tensor can be approximated by
\begin{eqnarray}
\label{TT}
T^t_t \sim T^{r}_{r}\sim-\frac{\eta^2}{r^2} \ , \ \ T^\theta_\theta=T^\phi_\phi\sim 0 \ .
\end{eqnarray}

As we can see global monopoles are spherically symmetric topological defects. They have been probably produced in early Universe due to a phase transition. As we shall see, asymptotically global monopoles present no Newtonian gravitational potential; however they give enormous tidal acceleration $a\approx 1/r^2$ which is important from the cosmological point of view, and may be used for obtaining upper limit on their number density present in the Universe today, which is at most one in the local group of galaxy \cite{H}. These objects have been investigated in the context of cosmology to explain the formation of large structure in the Universe \cite{VS}. Global monopoles present linear divergent energy strongly provided by the Higgs fields, $\varphi^a(x)$, which has to be cut-off at certain distance\footnote{The total energy inside a spherical region of radius $R$ is $E(R)\approx 4\pi G\eta^2R$. Considering $R$ the typical radius of a galaxy, $R=15 \ Kpc$ and for symmetry breaking scale $\eta\approx 10^{16}\ Gev$, which is the value for grand unified theories, this total energy is approximately ten times the mass of the galaxy.}. Moreover, global monopoles have also investigated in the context of brane-world. The monopole has its core at the $(3+1)-$dimensional brane and the Higgs fields in a $n=3$ extra dimensions \cite{Katherine,Inyong}.

Coupling the energy-momentum tensor associated with the global monopole system, (\ref{TT}), with the Einstein equations, solutions to the components of the metric tensor can be obtained. They are given below by the following line element:
\begin{equation}
ds^{2}=-\left( \alpha^{2}-\frac{2GM}r\right)dt^{2}+\left(\alpha^{2}-\frac{2GM}{r}\right)^{-1}dr^{2}+r^{2}d\Omega^{2} \ ,
\label{g}
\end{equation}
where $\alpha^{2}=1-8\pi G \eta^{2}<1$, and $M$ a constant of integration. For points very far from the monopole center, or simply discarding the integration constant, the above expression can be written by 
\begin{equation}
ds^{2}=-\alpha^{2}dt^{2}+\frac{dr^{2}}{\alpha^{2}}+r^{2}d\Omega^{2} \ .
\label{g1}
\end{equation}

Harari and Loust\'o \cite{HL}, presented a solution to the gravitational fields considering a simplified model to the monopole which displays the main features associated with this object. The authors model the monopole by a false vacuum in the region inside its core, i.e., $f=0$ for $r<\delta$, and by a true vacuum, $f=1$, in the region outside the monopole. So according to this model in the region inside, the solutions to the metric tensor is given by the following line element
\begin{equation}
ds^{2}=-\left( 1-H^{2}r^{2}\right)dt^{2}+\left( 1-H^{2}r^{2}\right)^{-1}dr^{2}+r^{2}d\Omega^{2} \ ,
\label{S}
\end{equation}
where
\begin{equation}
H^{2}=\frac{2\pi G \lambda \eta^{4}}{3} \ ,
\label{H}
\end{equation}
and in the region outside the monopole's core the metric tensor is given by (\ref{g}). Harari and Loust\'o also have pointed out that the inner de Sitter solution can match the exterior solution without the necessity to include an infinitely thin shell at the boundary, provided the mass, $M$, and the radius, $\delta$, of the monopole satisfy the conditions:
\begin{eqnarray}
\delta=\frac2{{\sqrt{\lambda}}\eta} \ , \ \ M=-\frac{16\pi}3\frac\eta{\sqrt{\lambda}} \ .
\end{eqnarray}
This negative value to the monopole mass does not violate the positive mass theorem because the spacetime is not asymptotically flat. So the gravitational mass does not coincide with the ADM mass \cite{U}.

The quantum analysis of the motions of massless fields in the spacetime described by the line element given by (\ref{g1}), have been developed for scalar \cite{ML} and fermionic \cite{Mello2} fields. In these publications the vacuum polarization effects due to these fields have been explicitly calculated. Specifically to the scalar field, Mazzitelli and Loust\'o have calculated the renormalized vacuum expectation value of the square of the field, $\langle\phi^2(x)\rangle_{Ren}$, and furnish, by dimensional analysis, the general structure of the renormalized vacuum expectation value of the energy-momentum tensor, $\langle T_{\mu\nu}(x)\rangle_{Ren}$. The authors have found that
\begin{eqnarray}
	\langle\phi^2(x)\rangle_{Ren}\approx \frac1{r^2} \ \quad{and} \ \ \ \langle T_{\mu\nu}(x)\rangle_{Ren}\approx \frac{S_{\mu\nu}}{r^4} \ .
\end{eqnarray}
So both polarization effects present divergences at $r=0$.  The origin of these singularities resides in considering the monopole as been a point-like object. Because it is not expected this kind of problem in a realistic model to a global monopole, a procedure that can be adopted to avoid it, is to consider the monopole as having a non-vanishing radius and consequently a inner  structure.

Motivated by this idea, we decided to return to the calculation of the vacuum polarization effects due to a massless scalar field in the global monopole spacetime considering at this time a non-vanishing radius to the latter. Moreover is also our objective estimate the effect of the monopole's core on the vacuum polarization in the region outside the monopole.
In order to develop this analysis we shall calculate the Euclidean Green function associated with this physical situation. Because the calculation of this Green function in the geometry described by (\ref{g}) cannot be written in a closed form, we shall consider the solution (\ref{g1}) for describing the exterior geometry. In this case, because the interior and exterior solutions cannot be smoothly matched, it necessary to include a surface layer at the boundary. In the thin wall approximation, the energy-momentum tensor associated with the layer is proportional to a delta-Dirac distribution concentrated on a surface. The analysis of the motion of a non static wall in the global monopole system has been analyzed a few years ago in \cite{GR,Cho}. In this present analysis we shall consider the wall being static. 

There are many methods to calculate the energy-momentum tensor associated with the infinitely thin wall \cite{D,I}. Here we shall adopt the distributional one \cite{MK}, which is equivalent to the presented by Darmois  \cite{D} and Israel \cite{I}. The interior and exterior metric to the wall must be continuous everywhere, however their derivative with respect the radial coordinate may be discontinuous. For this present case the thin wall must be a spherical surface of radius $r_0$ concentric with the monopole's core. So the metric tensor can be written in the following form:
\begin{equation}
g_{\mu \nu}=g_{\mu \nu}^{(+)}\Theta(r-r_0)+g_{\mu \nu}^{(-)}\Theta(r_0-r) \ ,
\label{g2}
\end{equation}
with $g_{\mu \nu}^{(+)}$ and $g_{\mu \nu}^{(-)}$ being the exterior and interior metric tensors, respectively, and $\Theta$ the step function. The energy-momentum tensor associated with the wall can be obtained by using the Einstein equation. We can express the energy-momentum tensor in whole space by:
\begin{equation}
T_{\mu \nu}(x)=T_{\mu \nu}^{(+)}(x)\Theta(r-r_0)+T_{\mu \nu}^{(-)}(x)\Theta(r_0-r)+T_{\mu \nu}^{(0)}(x)\delta(r-r_0) \ ,
\label{T}
\end{equation}
where the last term on the right hand side of the above equation corresponds to the contribution coming from the wall.  Calculating the Einstein tensor to the whole space by using ({\ref{g2}), we obtain that this tensor associated with the wall is given by terms proportional to delta-Dirac function, as shown bellow:
\begin{equation}
G^{\nu(0)}_\mu=r_0H^{2}diag(0,0,1,1)\delta(r-r_0) \ ,
\label{G0}
\end{equation}
being $r_0=\frac{2\sqrt{3}}{\eta \sqrt{\lambda}} $. So, using the Einstein equation we find
\begin{equation}
T_\mu^{\nu(0)}(x)=\frac{\sqrt{3 \lambda}}{6} \eta^{3}diag(0,0,1,1) \ .
\end{equation}
We can also write the Ricci scalar to the whole space 
\begin{equation}
{\cal{R}}(x)=12H^{2}\Theta(r_0-r)+\frac{2(1-\alpha^{2})}{r^{2}}\Theta(r-r_0)-\frac{16\pi G \sqrt{3\lambda}}6\eta^3\delta(r-r_0) \ .
\label{R}
\end{equation}

This paper is organized as follow. In section $2$ we explicitly calculate the Euclidean Green function associated with a massless scalar quantum field in the region outside the global monopole's core admitting a non-minimal coupling between the scalar field with the geometry: $\xi {\cal{R}}\hat{\Phi}^2$. We show that this Green function presents two parts: The first is the usual one associate with a point-like global monopole spacetime. The second is consequence of a nonvanishing structure admitted to the monopole's core. In section $3$ we calculate the renormalized vacuum expectation value $\langle\hat{\Phi}^2(x)\rangle$ in a complete form. We show that besides the standard result obtained to this expectation value, there appears a new contribution, the correction, consequence of assuming a non-vanishing value to the monopole's core. We also show that this quantity can be estimated in a approximate way. In general the correction decreases very fast with the distance to the monopole; however we show that for specific value of the non-minimal coupling constant, $\xi$, it presents a long-range effect. In section $4$ we present the formal expression to the renormalized vacuum expectation value of the energy-momentum tensor, $\langle\hat{T}_{\mu\nu}(x)\rangle$, and explicitly calculate the correction due to the inner structure of the monopole. We leave for section $5$ our conclusions and most relevant remarks about this paper.

\section{Calculation of the Euclidean Green Function}

The Euclidean Green function associated with a massless scalar field in the spacetime produced by a point-like global monopole, i.e., a spacetime whose geometry is described by the metric tensor given in (\ref{g1}) for the whole space, has been calculated by Mazzitelli and Loust\'o a few years ago \cite{ML}. There the Green function was obtained admitting a non-vanishing coupling between the field and the scalar curvature, ${\cal{R}}=2(1-\alpha^2)/r^2$.

In this section we want to improve this result calculating the Green function admitting a non-trivial structure to the monopole, and for points outside to its core. As we have already mentioned, our development will consider the spacetime described by two distinct metric tensor, given by (\ref{S}) for the region inside to the monopole and by (\ref{g1}) for the region outside to it. 

The Green function associated with a massless scalar field with a non-minimal coupling with the geometry must obey the non-homogeneous second-order differential equation
\begin{equation}
\left\{ \frac{1}{\sqrt{-g}}\partial_{\mu}\left[\sqrt{-g}g^{\mu\nu} \partial_{\nu}
\right]-\xi {\cal{R}}\right\}G(x,x^{'})=-\delta^{(4)}(x,x') \ .
\label{Green}
\end{equation} 

In order to calculate the vacuum average of the square of the scalar field and the energy-momentum tensor, let us construct the Euclidean version of the Green function by performing a Wick rotation on the time coordinate, i.e., $t\to i\tau$. Doing this and taking into account the components of the metric tensor considered in the regions inside (\ref{S}) and outside (\ref{g1}) of the monopole we get
\begin{equation}
\left\{ \frac{1}{h}\frac{\partial^{2}}{\partial r^{2}}+A(r)\frac{\partial}{\partial r}+\frac{1}{f}\frac{\partial^{2}}{\partial \tau^{2}}-L^{2}-\xi {\cal{R}}\right\}G(x,x^{'})=-\delta^{(4)}(x,x') \ ,
\label{Eq1}
\end{equation} 
where
\begin{equation}
f=-g_{00}=(1-H^{2}R^{2})\Theta(r_{0}-r)+\alpha^{2}\Theta(r-r_{0}) \ ,
\end{equation}
\begin{equation}
h=g_{11}=\frac{1}{(1-H^{2}R^{2})}\Theta(r_{0}-r)+\frac{1}{\alpha^{2}}\Theta(r-r_{0}) 
\end{equation}
and
\begin{equation}
A(r)=-\frac{1}{h^{2}}\left( \frac{dh}{dr}\right)+\frac{1}{2fh}\left( \frac{df}{dr}\right)+\frac{2}{rh} \ .
\end{equation}

Due to the spherical symmetry presented by this system, the Euclidean Green function can be written as
\begin{equation}
G(x,x')=\frac{1}{2\pi}\sum_{l=0}^{\infty}\sum_{m=-l}^{l}\int _{-\infty}^{\infty} dE e^{-iE(\tau-\tau')}g_{l}(r,r')Y^{\ast}_{lm}(\theta',\phi')Y_{lm}(\theta,\phi) \ .
\label{G1}
\end{equation}

Substituting (\ref{G1}) into (\ref{Eq1}) and using the standard representations to the $\delta-$function in coordinates $\tau$, $\phi$ and $\theta$, we arrive to the following differential equation for the unknown radial function $g_{l}(r,r')$:
\begin{equation}
\left[ \frac{1}{h} \frac{d^{2}}{dr^{2}}+A(r)\frac{d}{dr}-\frac{E^{2}}{f}-\frac{l(l+1)}{r^{2}}-\xi \mathcal{R}\right]g_{l}(r,r')=-\frac{1}{\sqrt{fh}r^{2}}\delta(r-r') \ .
\label{G2}
\end{equation}

All the informations about the inner structure of the monopole are contained in $g_l(r,r')$. This function must obey specific properties. Let us define by $g^<_l(r,r')$ the solution of (\ref{G2}) regular at $r\to 0$, and by $g^>_l(r,r')$ the solution that vanishes at infinity. These two functions satisfy the continuity condition at $r=r'$, with their first derivatives discontinuous at this point. 

Because it is our interest to calculate the vacuum polarization effect in the region outside the monopole's core, we have to solve (\ref{G2}) considering both $r$ and $r'$ greater than $r_0$. The solution to $g^<_l(r,r')$ must be continuous at $r=r_0$; however, because the curvature scalar possess a term proportional to $\delta(r-r_0)$, its first derivative is discontinuous at $r=r_0$.

So after these brief explanation about our procedure, let us now consider the solution of (\ref{G2}) with $r$ and $r'$ greater than $r_0$. In order to obtain the inner solution $g^<_l(r,r')$, we should consider two regions separately: from $0$ to $r_0$ and from $r_0$ to $r^{'}$. Integrating in the region $r<r'$, we find:
\begin{eqnarray}
g^<_l(r,r')=\left\{\begin{array}{cc}
A_{l}R_{l}(r),&\mbox{for $r<r_0$} \\
{B_{l}}\frac{1}{\sqrt{r}} I_{\nu_{l}}\left(\frac{E}{\alpha^{2}} r \right)+C_{l}\frac{1}{\sqrt{r}}K_{\nu_{l}}\left(\frac{E}{\alpha^{2}} r \right),
&\mbox{for $r_0<r<r'$,}
\end{array}
\right.
\label{g3}
\end{eqnarray}
where
\begin{equation}
R_{l}(r)=r^{l}{(H^{2}r^{2}-1)^{E/H}}_2F_1(\epsilon_l,\sigma_l,\beta_l;H^{2}r^{2}) \ ,
\end{equation}
being
\begin{equation}
\epsilon_l=\frac{E}{2H}+\frac{3}{4}+\frac{l}{2}-\frac{\sqrt{9-48\xi}}{4} \ ,
\end{equation}
\begin{equation}
\sigma_l=\frac{E}{2H}+\frac{3}{4}+\frac{l}{2}+\frac{\sqrt{9-48\xi}}{4} \ ,
\end{equation}
\begin{equation}
\beta_l=\frac32+l
\end{equation}
and
\begin{equation}
\nu_{l}=\frac{1}{\alpha}\sqrt{\left(l+\frac{1}{2} \right)^{2}+2(1-\alpha^{2})\left(\xi-\frac{1}{8} \right)} \ .
\end{equation}
In all the above expressions, $_2F_1$ represents the hypergeometric function. 

The outer solution of (\ref{G2}) is given by
\begin{equation}
g^>_l(r,r')=D_{l}\frac{1}{\sqrt{r}}K_{\nu_{l}}\left(\frac{E}{\alpha^{2}} r\right) \ , \quad \hbox{for $r>r'$.}
\label{g4}
\end{equation}

All the above constants are determined by imposing the boundary conditions obeyed by $g^<_l$ and $g^>_l$ at $r=r'$. Substituting the obtained results, we can write the function $g_l$ as shown below:
\begin{equation}
g_l(r,r')=\frac{1}{\alpha^{2}\sqrt{rr'}}K_{\nu_{l}}\left(\frac{E}{\alpha^{2}} r_{>} \right)\left[I_{\nu_{l}}\left(\frac{E}{\alpha^{2}} r_{<} \right)+\Omega_{l}(r_{0}) K_{\nu_{l}}\left(\frac{E}{\alpha^{2}} r_{<} \right) \right] \ ,
\label{g4}
\end{equation}   
where
\begin{equation}
\Omega_{l}(Er_0)=\frac{I'_{\nu_{l}}\left(\frac{E}{\alpha^{2}} r_{0} \right)R(r_{0})-I_{\nu_{l}}\left(\frac{E}{\alpha^{2}} r_{0} \right)\left[\left( \frac{1}{2r_{0}}+\frac{\xi \mathcal{R}^{(0)}}{\alpha^{2}}\right)R(r_{0})+R'(r_{0}) \right]}{K_{\nu_{l}}\left(\frac{E}{\alpha^{2}} r_{0} \right)\left[\left( \frac{1}{2r_{0}}+\frac{\xi\mathcal{R}^{(0)}}{\alpha^{2}}\right)R(r_{0})+R'(r_{0}) \right]-K'_{\nu_{l}}\left(\frac{E}{\alpha^{2}} r_{0} \right)R(r_{0})} \ ,
\label{O}
\end{equation}
being $\mathcal{R}^{(0)}=-\frac{8\pi G \sqrt{3\lambda}}3\eta^{3}$. In the above equations $r_>(r_<)$ is the larger (smaller) value between $r$ and $r'$, and the prime denotes the derivative with respect to the radial coordinates.

Substituting (\ref{g4}) into (\ref{G1}), and performing some intermediate steps, the Euclidean Green function is expressed by
\begin{eqnarray}
G(x,x')&=&\frac{1}{8\pi^{2}rr'}\sum_{l=0}^{\infty}(2l+1)Q_{\nu_{l}-\frac{1}{2}}(u_{o})P_{l}(\cos \gamma) \nonumber \\
&&+\frac{1}{4\pi \alpha^{2}\sqrt{rr'}}\sum_{l=0}^{\infty}(2l+1)P_{l}(\cos \gamma)\times \nonumber \\
&&\int_{0}^{\infty}dE\cos(E\Delta \tau)\Omega_{l}(Er_{0}) K_{\nu_{l}}\left(\frac{E}{\alpha^{2}} r_{>} \right)K_{\nu_{l}}\left(\frac{E}{\alpha^{2}} r_{<} \right) \ ,
\label{G3}
\end{eqnarray}
with 
\begin{equation}
u_0=\frac{r'^{2}+r^{2}+\alpha^{4}(\tau-\tau')^{2}}{2rr'} \ ,
\end{equation}
and $\gamma$ satisfying the the well known relation with the original angles $(\theta,\phi)$ and $(\theta',\phi')$. $Q_\lambda$ and $K_\lambda$ are the Legendre and modified Bessel functions, respectively. We can observe that the first term in (\ref{G3}) corresponds to the Green function obtained in \cite{ML} to the point-like global monopole spacetime, while the second one presents, besides a dependence on the $\alpha$ parameter, a dependence on the radius $r_0$ through the coefficient $\Omega_{l}$. 

Now  having the Green function we can calculate vacuum polarization effects of a massless scalar field in this spacetime.

\section{The Computation of $\langle\hat{\Phi}^{2}(x)\rangle_{Ren.}$}

In this section we calculate the renormalized vacuum expectation value of the square of the scalar field operator in the global monopole spacetime considering a inner structure to it. As we shall see this expectation value contains two contributions. The first one depends exclusively on the $\alpha$ parameter that we name {\it regular}, and the second, the correction, presents an explicit dependence on the radius of the monopole's core $r_0$. (The first contribution vanishes in the limit $\alpha\to 1$, while the second vanishes in the limit $r_0\to 0$.) 

The vacuum expectation value of the square of the scalar field is formally expressed by taking the coincidence limit of the Green function as shown below
\begin{equation}
\langle \hat{\Phi}^{2}(x)\rangle=\lim_{x'\rightarrow x}G(x,x') \ .
\label{P}
\end{equation} 

However the above procedure provides a divergent result which comes exclusively from the first term of (\ref{G3}) which is proportional to the behavior of the Legendre function evaluated at unity. So, in order to obtain a finite and well defined result to $\langle \hat{\Phi}^{2}(x)\rangle$, we must apply some regularization procedure. Here we adopt the renormalization procedure given in \cite{Wald}: we subtract from the complete Green function the Hadamard function. So the renormalized vacuum expectation value of the square of the scalar field is given by
\begin{equation}
\langle\hat{\Phi}^{2}(x)\rangle_{Ren.}=\lim_{x'\to x}\left[G(x,x')-
G_H(x,x')\right] \ ,
\label{P1}
\end{equation}
where $G_H(x,x')$ is the Hadamard function. Adopting this procedure and after some intermediate steps \cite{Grad} we find the following result:
\begin{eqnarray}
\langle\hat{\Phi}^{2}(x)\rangle_{Ren.}&=&\langle\hat{\Phi}^{2}(x)\rangle_{Reg.} \nonumber \\
&+&\frac{1}{4\pi \alpha^{2}r}\sum_{l=0}^{\infty}(2l+1)\int_{0}^{\infty}dE \Omega_{l}(Er_0)\left[ K_{\nu_{l}}\left(\frac{E}{\alpha^{2}} r \right)\right]^{2} \ .
\label{P2}
\end{eqnarray}

In their paper, Mazzitelli and Loust\'o \cite{ML} have computed, up to the first order in the parameter $\eta^2=1-\alpha^2$ considered smaller than unity, the first term of (\ref{P2}). The value obtained by these authors is
\begin{equation}
\label{phi2}
\langle\hat{\Phi}^{2}(x)\rangle_{Reg.}=-\frac{\eta^2}{4\pi^{2}r^{2}}\left[\frac{(p_0-2\xi p_1)}{2\sqrt{2}}+\left(\xi-\frac16\right)\ln(\mu r)\right] \ ,
\end{equation}
where $p_0 \approx -0.39$ and $p_1 \approx -1.41$. Because $\mu$ is completely arbitrary cutoff scale there is an ambiguity in the definition of the renormalized vacuum expectation value above. 

We can see that (\ref{phi2}) depends on the distance from the global monopole in a completely different manner compare to the case of an infinitely thin cosmic string spacetime due to the presence of the logarithmic term. The  second contribution of (\ref{P2}) is due to the nonvanishing value assumed to the monopole's radius. Allen {\it at al} \cite{Bruce} have calculated the vacuum polarization effect of a massless scalar field on a realistic cosmic string spacetime, considering generically the effect of the string's core through the non-minimal coupling between the scalar field with the geometry. There the authors found that the renormalized vacuum expectation value of the square of the field also presents two distinct contributions: the first being the standard one found in the literature, and the second, the correction, coming from the non-zero core radius of the string. Although being very small the radius of the cosmic string's core, the term of the correction associated with the $n=0$ component of the azimuthal quantum number, presents a long-range effect. This term decays very slowly with the distance to the cosmic string and can be approximated expressed by a term proportional to $1/r^2\ln(qr)$. 

Here the correction of (\ref{P2}) also cannot be expressed in a closed form. It is given in terms of an infinity sum of not solvable integrals. The complete information about this contribution can only be obtained numerically specifying the ration $r_0/r$, which we shall consider much smaller than unity. However, like in the paper \cite{Bruce}, an approximate behavior to the correction can be exhibited if we analyse the integrand of (\ref{P2}). Near the the origin $\Omega_l(Er_0)$ behaves as $(Er_0)^{2\nu_l}$, and $K_{\nu_l}(Er/\alpha^2)$ as $(Er)^{-\nu_l}$. Consequently the product 
\begin{eqnarray}
\Omega_l(Er_0)K^2_{\nu_l} (Er/\alpha^2)\approx (r_0/r)^{2\nu_l}<<1  \ . \nonumber
\end{eqnarray}
For $Er_0>>1$, $\Omega_l(Er_0)\approx e^{2Er_0}/{\sqrt{2Er_0}}$ and $K_{\nu_l}(Er/\alpha^2)\approx 
e^{-Er/\alpha^2}/{\sqrt{2Er/\alpha^2}}$. In this region we have
\begin{eqnarray}
\Omega_l(Er_0) K^2_{\nu_l}(Er/\alpha^2)\approx e^{-2Er/\alpha^2(1-\alpha^2 r_0/r)}\to 0 \ . \nonumber
\end{eqnarray}
All these informations indicate that we may approximate the integrand of (\ref{P2}) by assuming to the coefficient $\Omega_l(Er_0)$ its first-order expansion. Adopting this procedure we can write
\begin{equation}
\Omega_{l}(Er_0)=\frac{2}{\Gamma(\nu_{l}+1)\Gamma(\nu_{l})}\left[ \frac{\left(\nu_{l}-\frac{3}{4} \right)w_{l}-z_{l}}{\left(\nu_{l}+\frac{3}{4} \right)w_{l}+z_{l}} \right]\left( \frac{Er_{0}}{2 \alpha^{2}}\right)^{2 \nu_{l}} \ ,
\label{O1}
\end{equation}
where $w_l$ and $z_l$ are given by
\begin{equation}
w_l={_2F_1}(\bar{\epsilon}_l,\bar{\sigma}_l,\beta_l;1-\alpha^2)
\end{equation}
and
\begin{equation}
z_l=l_2F_1(\bar{\epsilon}_l,\bar{\sigma}_l,\beta_l;1-\alpha^2)+\frac{2\bar{\sigma} \bar{\epsilon}_l(1-\alpha^2)}{\beta_l}{_2F_1}(\bar{\epsilon}_l+1,\bar{\sigma}_l+1,\beta_l+1;1-\alpha^{2}) \ ,
\end{equation} 
with 
\begin{equation}
\epsilon_l\approx \frac{3}{4}+\frac{l}{2}-\frac{\sqrt{9-48\xi}}{4} \ ,
\end{equation}
\begin{equation}
\sigma_l\approx \frac{3}{4}+\frac{l}{2}+\frac{\sqrt{9-48\xi}}{4} \ .
\end{equation}
Consequently an approximate expression to (\ref{P2}) can be obtained:
\begin{eqnarray}
\langle\hat{\Phi}^{2}(x)\rangle_{Ren.}&=&\langle\hat{\Phi}^{2}(x)\rangle_{Reg.}+\frac{1}{4r^2}
\sum_{l=0}^{\infty}\frac{(2l+1)}{2^{\nu_l}\Gamma(\nu_l)[\Gamma(1+\nu_l/2)]^2}\times\nonumber\\
&&\left[\frac{(\nu_l-3/4)w_l-z_l}{(\nu_l+3/4)w_l+z_l}\right]\Gamma\left(\frac{1+3\nu_l}2)\right)\Gamma\left(\frac{1-\nu_l}2\right)\left(\frac{r_0}r\right)^{\nu_l} \ .
\end{eqnarray}

Because $r_0/r$ is assumed to be smaller than unity, the most relevant correction comes from the $l=0$ component. In this case the extra radial dependence of the second term is dominated by $(r_0/r)^{\frac1{2\alpha}}$ for $\xi=1/8$. 

The most  significant change in the behavior of the correction of (\ref{P2}), happens for the case where the order of the modified Bessel function vanishes. In principle there is no physical evidence that determine the value of the non-minimal coupling constant $\xi$. However, for specific value of this constant, there appears in this model a long-range effect on the correction. The specific value will be given by imposing $\nu_l=0$. So for the $l=0$ component this value is  $\xi=\xi_0= \frac18(1-(1-\alpha^2)^{-1})$. \footnote{We may adopt a different value of $\xi$ which vanishes $\nu_l$ for $l=n>0$, being $n$ a positive integer number. However this value will produce complex value for $\nu_l$ for $l<n$.} Accepting this special value for the coupling constant the behavior of the coefficient $\Omega_0$ for $Er_0$ smaller than unity changes completely. It will be given by
\begin{equation}
\Omega_0(Er_0)=\frac1{\ln\left(\frac{Er_0}{2\alpha^2}\right)+\mathcal{C}'} \ ,
\label{O2}
\end{equation}
where $\mathcal{C}'=\mathcal{C}-w_0\left(z_0+3w_0/4\right)^{-1}$ being $\mathcal{C}$ the Euler constant. The approximate values to $f_0$ and $g_0$ are obtained from the general expressions given before taking $l=0$. In this way we have
\begin{equation}
\bar{\epsilon}_0\approx\frac34-\frac14\sqrt{\frac{9-3\alpha^2}{1-\alpha^2}}
\end{equation}
and
\begin{equation}
\bar{\sigma}_0\approx\frac34+\frac14\sqrt{\frac{9-3\alpha^2}{1-\alpha^2}} \ .
\end{equation}

As we can see for $l \not= 0$, $\Omega_{l}(Er_{0}/\alpha^{2})$ decreases at least with $(Er_{0}/\alpha^{2})^{2\nu_{l}}$ for $(Er_{0}/\alpha^{2}) \rightarrow 0$. So their respective contributions to $\langle\hat{\Phi}^{2}(x)\rangle_{Ren.}$ behave as $(r_0/r)^{\nu_l}$.  On the other hand $\Omega_{0}(Er_{0}/\alpha^{2})$ decreases slowly with the inverse of the logarithm in this limit. Consequently the most relevant contribution to the sum in (\ref{P2}) comes from the $l=0$ component. This means that we can substitute the whole sum by just one term \footnote{This result has been confirmed numerically adopting $\alpha =0.9$ and $r_{0}/r=10^{-3}$.}. Let us use a more convenient notation to express (\ref{O2})
\begin{equation}
\Omega_{0}=\frac{1}{\ln\left(\frac{E}{\alpha^{2} q}\right)} \ ,
\label{O3}
\end{equation}
with
\begin{equation}
q=\frac{2}{r_{0}}e^{-\mathcal{C}'} \ .
\label{q}
\end{equation}

The analysis of the behavior of the integrand of (\ref{P2}) assuming the expression (\ref{O3}) to $\Omega_0$ can also be developed in a formal way as follows: in the region $Er$ much smaller than unity, $K_0(Er/\alpha^2)$ behaves as $-\ln(Er/\alpha^2)$. So the product
\begin{equation}
\Omega_0(Er_0)\left[K_{0}\left(\frac{E}{\alpha^{2}}r\right)\right]^{2}\approx \frac{\left[\ln \left(\frac{E}{\alpha^{2}}r\right) \right]^2}{\ln\left(\frac{E}{\alpha^{2} q}\right)}\approx \ln\left(\frac{E}{\alpha^{2}}r\right) \ .
\end{equation}
Although being divergent the above expression, it can be integrated in the neighborhood $Er\approx 0$. For $Er$ much bigger than unity, $\Omega_0(Er_0)\approx e^{2Er_0}$ and $K_0(Er/\alpha^2)\approx e^{-Er/\alpha^2}$. Consequently the product $\Omega_0K^2_0$ goes to zero in this limit.

So the most relevant contribution of the correction to the renormalized vacuum expectation value of the square of the operator scalar field is proportional to
\begin{equation}
\int_0^{\infty} dE \frac{K^2_0(\frac{E}{\alpha^{2}} r)}{\ln\left(\frac{E}{\alpha^{2}q} \right)}=\frac{\alpha^{2}}{r}
\int_0^{\infty} dv  \frac{K^2_0(v)}{\ln(v)-\ln(qr)} \ .
\label{27}
\end{equation}

As it has been discussed Bruce {\it{et al}} \cite{Bruce}, this approximation presents a problem because of the integrand has a pole at $v=qr$. However, this pole is a consequence of the approximation adopted, and it occurs in the region where the argument of $\Omega_0$ is bigger than unity, consequently the approximation is no longer valid. Here, in our calculation, we reached the same conclusion. Choosing $\alpha=0.9$, the value assumed for $\mathcal{C}'$ is $0.68$. So $qr=\frac{2 r}{r_{0}}e^{-{\mathcal{C}}'}$ is of order $10^{3}$ for $r_{0}/r=10^{-3}$ and $Er_{0}/\alpha^{2} \approx 1$. Besides we can observe numerically that (\ref{27}) can be approximated discarding the term $\ln(v)$. 

Supported by this analysis (\ref{27}) can be written as
\begin{equation}
-\frac{\alpha^2}r\frac1{\ln(qr)}\int_0^\infty \ dv\ K_0^2(v) \ ,
\label{Y}
\end{equation}
consequently  we conclude that the second term of (\ref{P2}) can be approximate by \cite{Grad}
\begin{equation}
\label{X}
-\frac\pi{16} \frac1{r^{2}\ln(qr)} \ .
\end{equation}

As we have already discussed our approximation consists to substitute the expression (\ref{O2}) to the coefficient $\Omega_0$ in (\ref{P2}) and discard the pole. In order to confirm this approximation, we analyzed numerically the integrand of the second contribution of (\ref{P2}) using the exact expression (\ref{O}) and the approximate one (\ref{X}) to the correction. The numerical results are exhibit in Fig. 1. 

We finally present the expression below to the renormalized vacuum expectation value of the square of the field operator in the global monopole spacetime considering a inner structure to it:
\begin{eqnarray}
\langle\hat{\Phi}^{2}(x)\rangle_{Ren.}&=&-\frac{\eta^2}{4\pi^{2}r^{2}}\left[\frac{(p_{0}-2\xi_{o} p_{1})}{2\sqrt{2}}+\left(\xi_{o}-\frac{1}{6}\right)\ln(\mu r)\right] \nonumber \\
&&-\frac{\pi}{16 r^{2}\ln \left(\frac{2 r}{r_{0}}e^{-\mathcal{C}'} \right)} \ .
\end{eqnarray}

As we can see the explicit dependence on the inner structure of the monopole appears only in the third term. This term vanishes slowly with $r\rightarrow \infty$, consequently it is a long-range effect.
\begin{figure}[!htb]
\begin{center}
\includegraphics[width=7cm,angle=-90]{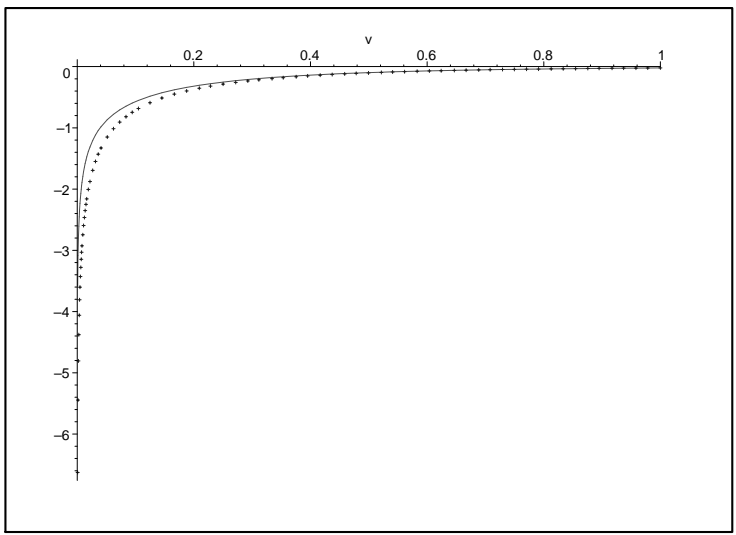}
\label{fig1}\caption{The solid curve represents the integrand of equation (\ref{P2}) written in terms of the new variable $v=Er/\alpha^{2}$ considering the complete expression for the coefficient $\Omega_0$ given by (\ref{O}). The dashed curve represents the approximate integrand given by (\ref{Y}) including the prefactor. In this numerical analysis we have used $\alpha=0.9$ and $r_{0}/r=10^{-3}$. }
\end{center}
\end{figure}

\section{Computation of $\langle \hat{T}_{\mu \nu}\rangle _{Ren}$}

The energy-momentum tensor $T_{\mu \nu}(x)$ is a bilinear function of the fields, consequently we can calculate its vacuum expectation value, $\langle T_{\mu \nu}(x)\rangle$, by standard method using the Green function \cite{Birrel}. 

The general structure of the renormalized vacuum expectation value of the energy-momentum tensor associated with a massless scalar field in a point-like global monopole spacetime has be developed by Mazzitelli and Lousto \cite{ML} a few years ago. There the authors have shown that this tensor can be written as
\begin{equation}
\langle\hat{T}_{\mu \nu}(x)\rangle_{Reg}=\frac1{16\pi^2r^4}\left[A_{\mu\nu}+B_{\mu\nu}\ln(\mu r)\right] \ .
\end{equation}
Although the exact expression to the tensor $A_{\mu\nu}$ has not been been explicitly calculated yet, its general structure can be provided by imposing that $\langle\hat{T}_{\mu \nu}(x)\rangle_{Reg}$ must be conserved
\begin{equation}
\nabla_\mu \langle \hat{T}_\nu^\mu \rangle_{Reg}=0 
\end{equation}
and satisfies the conformal trace anomaly
\begin{equation}
\langle \hat{T}_\mu^\mu\rangle _{Reg}= \frac1{16\pi^2}a_2(x)
\end{equation} 
for $\xi=\frac16$.

Because the Green function associated with this system presents two distinct terms, $\langle \hat{T}_{\mu \nu}\rangle_{Ren} $ will be expressed by two contributions:
\begin{equation}
\langle\hat{T}_{\mu\nu}\rangle_{Ren}=\langle\hat{T}_{\mu \nu}\rangle_{Reg}+\langle \hat{T}_{\mu \nu}\rangle_{C} \ ,
\end{equation}
where $\langle \hat{T}_{\mu \nu}\rangle _{Reg}$ is the contribution coming from the point-like global monopole spacetime, and $\langle \hat{T}_{\mu \nu}\rangle_{C}$ the contribution coming from the non-vanishing core radius of the monopole. Here we present the correction to this expectation value, which can be calculated by using the expression below
\begin{eqnarray}
\langle\hat{T}_{\mu \nu}\rangle_C&=&\lim_{x'\rightarrow  x}\left[(1-2\xi)\nabla_\mu\nabla_{\nu'}G_C(x,x')-2\xi\nabla_\mu\nabla_\nu G_{C}(x,x')\right. \nonumber \\
&&+\left.\left(\xi-\frac14\right)g_{\mu\nu}(x)g^{\rho \sigma'}(x,x')\nabla_\rho\nabla_{\sigma'}G_{C}(x,x')\right. \nonumber \\
&&\left.-\xi G_{\mu \nu}(x)G_C(x,x')-2\xi^2g_{\mu\nu}\mathcal{R}(x)G_C(x,x')\right] \ ,
\label{TC}
\end{eqnarray}
being $G_{\mu \nu}$  the Einstein tensor and $G_{C}(x,x')$ the correction of the Green function. Because the second contribution to the Green function is finite at the coincidence limit, the above calculation can be performed directly.

Supported by our previous discussion we can infer that the most important contribution to $\langle \hat{T}_{\mu \nu}\rangle_C$ comes from the $l=0$ component of (\ref{G3}). Considering only the specific situation where the order of the modified Bessel function vanishes, we found the following results:
\begin{eqnarray}
\langle \hat{T}_{00}\rangle_{C}&=&\frac{1}{4\pi r^{2}}\left(\xi_{0}-\frac{1}{4}\right)\int_{0}^{\infty}dEE\Omega_{0}(Er_{0})K_{0}\left(\frac{Er}{\alpha^{2}}\right)K_{1}\left( \frac{Er}{\alpha^{2}}\right) \nonumber \\
&&+\frac{1}{4\pi r^{3}}\left(\xi_{0}-\frac{1}{4}\right)\left[\frac{1}{16}-\xi_{0}(1-\alpha^{2})\right]\int_{0}^{\infty} dE\Omega_{0}(Er_{0})K^{2}_{0}\left( \frac{Er}{\alpha^{2}}\right) \nonumber \\
&&+\frac{1}{4\pi \alpha^{2}r}\int_{0}^{\infty} dEE^{2}\Omega_{0}(Er_{0})\times \nonumber \\
&&\left[\left(\frac{3}{4}+\xi_{0}\right)K^{2}_{0}\left(\frac{Er}{\alpha^{2}}\right)+\left(\xi_{0}-\frac{1}{4}\right)K^{2}_{1}\left( \frac{Er}{\alpha^{2}}\right) \right] \ ,
\label{t0}
\end{eqnarray}
\begin{eqnarray}
\langle \hat{T}_{11}\rangle_{C}&=&\frac{1}{16\pi \alpha^{6}r}\int_{0}^{\infty}dEE^{2}\Omega_{0}(Er_{0})\times \nonumber \\
&&\left[ (4\xi_{0}-3)K^{2}_{1}\left( \frac{Er}{\alpha^{2}}\right)-(4\xi_{0}+1)K^{2}_{0}\left( \frac{Er}{\alpha^{2}}\right)\right] \nonumber \\
&&+\frac{1}{16\pi \alpha^{4}r^{2}}\left( 3-20\xi_{0} \right)\int_{0}^{\infty}dEE\Omega_{0}(Er_{0})K_{0}\left( \frac{Er}{\alpha^{2}}\right)K_{1}\left( \frac{Er}{\alpha^{2}}\right) \nonumber \\
&&+\frac{1}{\pi \alpha^{2}r^{3}}\left[ \frac{3}{64}-\frac{7\xi_{0}}{16}+\xi_{0}(4\xi_{0}-1)\left(1-\frac{1}{\alpha^{2}} \right)\right]\times \nonumber \\
&&\int_{0}^{\infty} dE\Omega_{0}(Er_{0})K^{2}_{0}\left( \frac{Er}{\alpha^{2}}\right) 
\label{t1}
\end{eqnarray}
and
\begin{eqnarray}
\langle \hat{T}_{22}\rangle_{C}=\frac{1}{\sin^{2}(\theta)}\langle \hat{T}_{33}\rangle_{C}=-\frac{\xi_{0}^{2}(1-\alpha^{2})}{\pi \alpha^{2}r}\int_{0}^{\infty} dE\Omega_{0}(Er_{0})K^{2}_{0}\left( \frac{Er}{\alpha^{2}}\right) \ .
\label{t3}
\end{eqnarray}

Finally substituting (\ref{O3}) into (\ref{t0})-(\ref{t3}), and taking into account the previous discussion to evaluate the integral in the calculation of the vacuum expectation value of the square of the scalar field, we obtain:
\begin{eqnarray}
\langle\hat{T}_\mu^\nu(x)\rangle_{C}=\frac{\pi\alpha^{2}}{512(1-\alpha^{2})\ln(qr)r^{4}}diag(10-3\alpha^{2},8\alpha^{2}-4,2\alpha^{2},2\alpha^{2}) \ .
\end{eqnarray}

As we can see, the above contribution decreases slowly with the distance to the monopole's core with $1/r^4\ln(qr)$. Because in our approximation we have discarded all the other components, it is not surprise that $\langle \hat{T}_{\mu \nu}\rangle_C$ violates the conservation condition. In fact we have shown that this violation vanishes for $r_0\to 0$.

\section{Concluding Remarks}

Previous calculations of the vacuum polarization effects on spacetimes produced by global monopoles present divergences at the monopoles' position. The origin for these singularities resides in the fact of these objects have been considered as point-like ones. Realistic models to global monopoles necessarily have to consider a inner structure to them. Unfortunately it is not possible to provide the exact behavior of their structure, only numerical analysis ca do that. 

The main objective of this paper is to calculate the vacuum polarization effects due to a massless scalar field in the global monopole spacetime considering a inner structure to it. The basic motivation was try to avoid the singularities present in the vacuum average of the square of the field and also in the energy-momentum tensor at origin. Specifically, in this paper we have adopted the following approximated model to the global monopole spacetime: we have considered the spacetime as been de Sitter in the region inside the monopole's core, and point-like in the region outside. Moreover, our calculations have been developed for points outside the monopole's core. For this model we have shown that the renormalized vacuum expectation value of the square of the scalar field presents two distinct contributions: the first is the usual one found in the literature associated with a point-like object, and the second, the correction, is due to its inner structure. We have pointed out that for specific value of the non-minimal coupling constant $\xi$, the correction presents a long-range effect. We also calculated the correction to the renormalized vacuum expectation value of the energy-momentum tensor due to the inner structure to the monopole. As in the previous analysis, we show that a long-range effect can also takes place on this quantity. As we have seen, for both calculations, $\langle\hat{\Phi}(x)^2\rangle_C$ and $\langle\hat{T} _\mu^\nu(x)\rangle_{C}$, the extra radial dependences presented by these quantities are equal, and given by $1/\ln(qr)$, which is smaller than any power of the inverse of $r$. 

In \cite{Bruce}, Bruce {\it{et al}} have calculated the vacuum polarization effect due to a massless scalar field in the spacetime produced by a cosmic string considering a general structure to it. The author have observe that besides the standard contribution, there appears a correction due to the non-zero core radius of the string. They have also shown that the correction has the same magnitude as the geometrical contribution up to a distance from the cosmic string that exceed the radius of the observable universe. 

If we are inclined to make the same kind of comparison in the global monopole spacetime, between the standard contribution to vacuum polarization effect with the one coming from the correction, we should take into consideration that, because this spacetime has a non-vanishing scalar curvature, there appears in these vacuum averages an arbitrary cut-off energy scale $\mu$. Consequently considering the case with specific value to the non-minimal coupling constant $\xi$, the radial distance where both effects have the same magnitude should depend on $\mu$.

\newpage

{\bf{Acknowledgment}}
\\       \\
One of us (ERBM) wants to thanks Conselho Nacional de Desenvolvimento Cient\'\i fico e Tecnol\'ogico (CNPq.), and also CNPq./FAPESQ-PB (PRONEX), for partial financial support.


\begin{thebibliography}{9}

\bibitem{BV} M. Barriola e A. Vilenkin, {\it Phys. Rev. Lett.} {\bf 63}, 341 (1989). 
\bibitem{All} Dieter Maison and Steven L. Liebling, {\it Pys. Rev. Lett.} {\bf 25}, 5218 (1999); Steven L. Liebling, {\it Phys. Rev.} D {\bf 60}, 061502 (1999); Ulises Nucamendi, Marcelo Salgado and Daniel Sudarski, {\it Pys. Rev. Lett.} {\bf 14}, 3037 (2000).
\bibitem{Mello1} J. Spinelly, U de Freitas and E. R. Bezerra de Mello, {\it Phys. Rev.} D {\bf 66}, 024018 (2002); Yves Brihaye and Betti Hartmann, {\it Phys. Rev.} D {\bf 66}, 064018 (2002); Eug\^enio R. Bezerra de Mello, Yves Brihaye and Betti Hartmann, {\it Phys. Rev.} D {\bf 67}, 045015 (2003).  
\bibitem{H} W. Hiscock, {\it Phys. Rev. Lett.} {\bf 64}, 344 (1990).
\bibitem{VS} A. Vilenkin and E. P. S. Shellard, {\it Cosmic Strings and Other Topological Defects}, Cambridge University Press, Cambridge (1994).
\bibitem{Katherine} Katherine Benson and Inyong Cho, {\it Phys. Rev.} D {\bf 64}, 065026 (2001).
\bibitem{Inyong} Inyong Cho, {\it Phys. Rev.} D {\bf 68}, 025013 (2003).
\bibitem{HL} D. Harari e C. Loust\'{o}, {\it Phys. Rev.} D {\bf 42}, 2626 
(1990).
\bibitem{U} U. Nucamendi and D. Sudarsky, {\it Class. Quantum Gravity} {\bf 14}, 1309 (1997).
\bibitem{ML} F. D. Mazzitelli and C. O. Loust\'o, {\it Phys. Rev.} D  {\bf 43}, 468 (1991). 
\bibitem{Mello2} E. R. Bezerra de Mello, V. B. Bezerra and N. R. Khusnutdinov, {\it Phys. Rev.} D {\bf 60}, 063506 (1999).
\bibitem{GR} E. I. Guendelman and A. Rabinowitz, {\it Phys. Rev.} D {\bf 44}, 3152 (1991).
\bibitem{Cho} Inyong Cho and Jemal Guven, {\it Phys. Rev.} D {\bf 58}, 063502 (1998).  
\bibitem{D} G. Darmois, {\it Les Equation de la Gravitation Einsteinnienne}, in {\bf Memorial des Science Mathematiques}, Fasc. XXV, Chap. V (1927).
\bibitem{I} W. Israel, {\it Nuovo Cimento} B {\bf 44}, 1 (66); W. Israel, ibid. {\bf 48}, 463 (1967).
\bibitem{MK} R. Mansouri and M. Khorrami, {\it J. Math. Phys.} {\bf 37}, 5672 (1996).

\bibitem{Wald} R. Wald, {\it Commu. Math. Phys.} {\bf 54}, 1 (1977).
\bibitem{Birrel}  N. D. Birrell and P. C. W. Davies. {\it Quantum Fields
in Curved Space} (Cambridge University Press. Cambridge. England. 1982).
\bibitem{Grad} I. S. Gradshteyn and I. M Ryzhik, {\it{Table of Integral, Series and Products}} (Academic Press, New York, 1980). 
\bibitem{Bruce}B. Allen, B. S. Kay and A. C. Ottewill, {\it Phys. Rev.} D {\bf 53}, 6829 (1996).
 
\end{thebibliography}
\end{document}